\documentclass[journal=jacsat,manuscript=article]{achemso}

\usepackage[version=3]{mhchem} 

\author{Sviatoslav Kovalchuk}
\altaffiliation{Contributed equally to this work}
\affiliation{Department of Physics, Freie Universit\"{a}t Berlin, 14195 Berlin, Germany}
\author{Moshe G. Harats}
\altaffiliation{Contributed equally to this work}
\affiliation{Department of Physics, Freie Universit\"{a}t Berlin, 14195 Berlin, Germany}
\author{Guillermo L\'{o}pez-Pol\'{i}n}
\affiliation{Department of Physics, Freie Universit\"{a}t Berlin, 14195 Berlin, Germany}
\author{Jan N. Kirchhof}
\affiliation{Department of Physics, Freie Universit\"{a}t Berlin, 14195 Berlin, Germany}
\author{Katja Höflich}
\affiliation{Helmholtz Zentrum für Materialien und Energie Berlin, Hahn-Meitner-Platz 1, D — 14109 Berlin, Germany}
\author{Kirill I. Bolotin}
\affiliation{Department of Physics, Freie Universit\"{a}t Berlin, 14195 Berlin, Germany}
\email{bolotin@zedat.fu-berlin.de}

\title[An \textsf{achemso} demo]
  {Neutral and charged excitons interplay in non-uniformly strain-engineered WS$_2$}

\abbreviations{TMDC, PL, $WS_2$}

\begin{document}


\begin{abstract}
We investigate the response of excitons in two-dimensional semiconductors subjected to controlled non-uniform strain fields. In our approach to non-uniform strain-engineering, a WS$_2$ monolayer is suspended over a triangular hole. Large ($>2\;\%$), strongly non-uniform ($>0.28\;\%/\mu m$), and in-situ tunable strain is induced in the monolayer by pressurizing it with inert gas. We observe peak shifts and spectral shape changes in the photoluminescence spectra of strained WS$_2$. We interpret these changes as a signature of increased free electron density and resulting conversion of neutral excitons to trions in the region of high strain. Our result establishes non-uniform strain engineering as a novel and useful experimental `knob' for tuning optoelectronic properties of 2D semiconductors.
\end{abstract}

\section{Introduction}

Mechanical strain is a convenient physical `knob' that allows manipulation of various material properties. Strain engineering has been used, for example, to improve carrier mobility in Silicon transistors\citep{strain_mobility, strain_mobility1, strain_mobility2, strain_mobility3, strain_mobility4} and to reduce the switching energy in phase change memories\citep{phase_change_memory}. Two-dimensional materials such as monolayer transition metal dichalcogenides (TMDCs) are especially amenable to strain engineering. These materials are thin, flexible, and very stiff. Specifically, the breaking strain of TMDCs is higher than 10 \%\citep{breakingstrain15, breakingstrain16, breakingstrain17, breakingstrain}, while Young's modulus reaches $250\;GPa$ for WS$_2$\citep{breakingstrain15}. This leads to a large tunability of the size and the nature (direct vs. indirect) of the band-gap\citep{bandgap19, strain_pl6, indirect_direct}, phonon frequencies\citep{detuning}, and scattering mechanism of TMDCs\citep{tmdc_raman, bunch} via strain engineering. Moreover, strain distributions in TMDCs can be simply visualized via photoluminescence (PL) spectroscopy measurements\citep{bunch, strain_pl1, strain_pl2, strain_pl3, strain_pl4, strain_pl5, strain_pl6, detuning}.

While the effects of uniform strain have been studied extensively\citep{strain_pl4, strain_pl5, strain_pl6, detuning}, spatially non-uniform strain received much less attention. Although challenging to realize experimentally, non-uniform strain gives rise to theoretically predicted effects including exciton funneling\citep{funneling}, pseudomagnetic fields\citep{pseudo_magnetic, pseudo_magnetic2}, and formation of quantum-dot like states\citep{single_photon}. Very recently, we realized a powerful new scheme for non-uniform strain engineering in a prototypical TMDC tungsten disulphide (WS$_2$)\citep{moshe}. The strain was induced by a sharp AFM tip indenting a WS$_2$ membrane and the strain effects were probed via PL spectroscopy. In these experiments, we found signatures of a new effect – strain related exciton-to-trion conversion. However, the AFM-based technique has its shortcomings. While steep strain gradients are induced in the proximity of the AFM tip, the strain falls rapidly with distance from the tip and largely relaxes a micron away. This makes it difficult to map the strain by optical methods that are limited in resolution by the diffraction limit. Unavoidable drifts in the setup led to rupturing of membranes over hour-long observation periods. In addition, the technique is relatively complex and requires a custom AFM setup. This precludes the use of non-uniformly strain engineered samples in other experimental setups.

Here, we develop another, much simpler approach to non-uniform strain engineering.
TMDC membrane suspended over a hole is strained by pressurizing gas in contact with it
Critically, by choosing the geometry of the hole (i.e. circular or triangular), we define the specific strain distribution and the degree of its non-uniformity. By maintaining the pressure difference across the membrane, the TMDC is kept in a constant strain state for hours. This enables us to perform full spectroscopic mapping of TMDC monolayers subjected to non-uniform strain. We use such mapping as well as comparison with Finite Element Method (FEM) simulations to directly visualize large, $>2\;\%$, and highly non-uniform, $>0.28\;\%/\mu$m (at $1.7\;bar$ pressure), strain induced in our devices. Other advantages of our technique are simplicity and compactness of the setup, easy access to the sample, and the predictability of the resulting strain distribution. We study the effects of non-uniform strain and confirm that they qualitatively differ from the previously studied and well-understood effects of uniform strain. Specifically, we demonstrate conversion of neutral excitons to trions in the highly-strained region of the membrane. We show that this is the result of  strain-related `funneling' of free carriers into that region. While pressurized membranes are widely used for induction and investigation of uniform strain in 2D materials \citep{bunch, pressure1, pressure3}, this is the first study, to the best of our knowledge, targeting controlled non-uniform strain in such membranes.


\begin{figure}[H]
  \includegraphics[width=\textwidth]{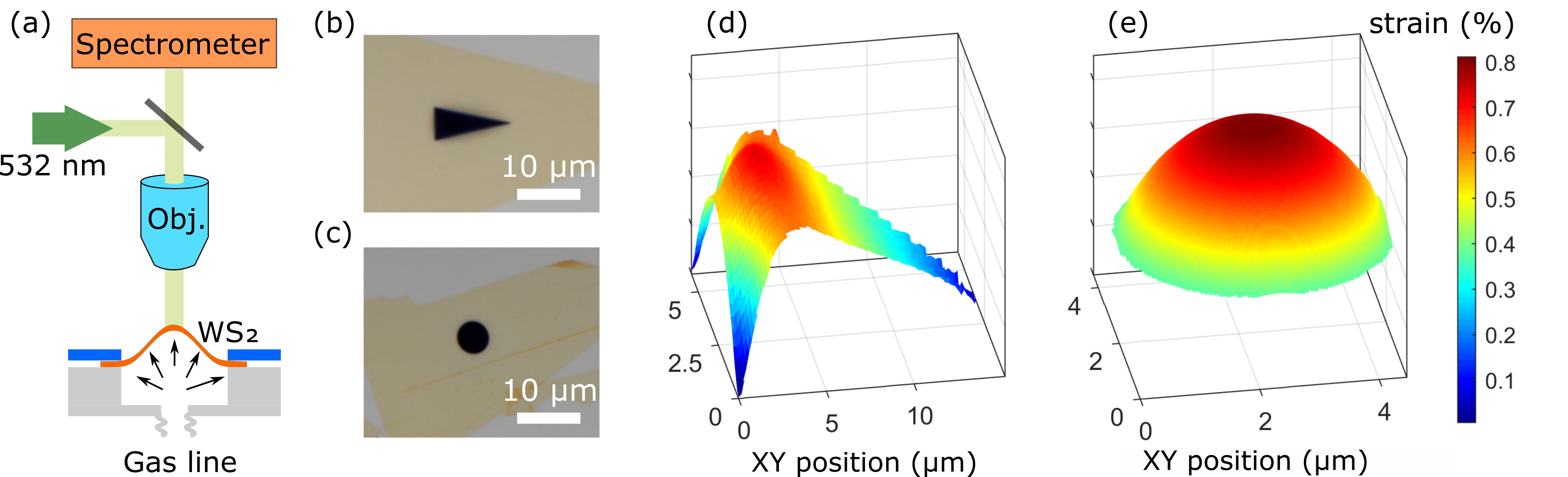}
  \caption{(a) Measurement setup schematics. The WS$_2$ flake suspended over a hole in a SiNx membrane is pressurized from one side and optically interrogated from another. (b-c) Optical images of monolayer WS$_2$ flakes transferred over a triangular (b) and a circular (c) hole. (d-e) The spatial variation of the local strain $\varepsilon_{xx}(x,y)+\varepsilon_{yy}(x,y)$ simulated by a finite element method (FEM) simulation for a triangular (d) and circular (e) membranes, respectively. Pressure difference is $\Delta P=0.8\;bar$ for both geometries. The colorbar presents the strain magnitude.}
  \label{fig:one}
\end{figure}

\section{Setup}
We chose a prototypical TMDC monolayer WS$_2$ to explore the effects of non-uniform strain-engineering. This material is characterized by bright PL\citep{ws2lum} and well-resolved excitonic peaks compared to other TMDCs. The samples consist of WS$_2$ monolayers transferred over a hole in a SiNx membrane (Fig. \ref{fig:one}b,c). We use commercial semi-transparent membranes with SiNx windows $20\;nm$ thick and $20\times20\;\mu m$ in size (Norcada). To investigate different strain fields, we use circular ($5\;\mu m$ in diameter) and triangular ($12\;\mu m$ side and $5\;\mu m$ base lengths) holes in SiNx (Fig. \ref{fig:one}b,c). Prior to the experiments, we simulated (Comsol\texttrademark) the distribution of strain in both experimental geometries (Fig. \ref{fig:one}d,e). By comparing Fig. \ref{fig:one}d and Fig. \ref{fig:one}e, we see that strain is more non-uniform in the case of triangular holes. The holes in SiNx are cut using a focused beam of Gallium ions (Crossbeam $340$ FIB) at $30\;keV$ acceleration voltage and a current of $50\;pA$. Afterwards, $100\;nm$ of gold is deposited onto the SiNx substrate to increase the stiffness of the SiNx layer and to avoid spurious deformation. Next, WS$_2$ monolayers are mechanically exfoliated onto an intermediary PDMS substrate and characterized using PL. Finally, suitable WS$_2$ monolayers are mechanically transferred onto holes in SiNx using a well-established dry-transfer technique\citep{exfoliation}.

To induce strain in the suspended WS$_2$, the sample is mounted onto a miniature pressure chamber (Fig. \ref{fig:one}a). The sample is placed with the WS$_2$ monolayer facing down to avoid unwanted slippage between the WS$_2$ and the SiNx membrane, and glued to the chamber using Pattex Natural 26 epoxy. Finally, inert gas at controlled pressure regulated by a manometer with an accuracy of $1\;mbar$ is connected to the chamber, thereby pressurizing the monolayer. We have made measurements with $N_2$ and $He$ gases, to check for possible effects of gas-related doping\citep{vacancies}, but detected no such effects (see Supporting Information Fig. S1). We report the pressure magnitude in terms of $\Delta P$, the difference between the pressure inside the chamber and the atmospheric pressure outside.

PL maps of strain-engineered WS$_2$ are acquired using a periscope shown in Fig. \ref{fig:one}a. We use a high resolution objective with $NA=0.75$ and precise translation stages to reach sub-micron mapping precision. The sample is excited with a CW laser emitting at $532\;nm$. The beam is tightly focused (FWHM $600\;nm$) with a constant power of $15\;\mu W$ unless stated otherwise. PL spectra WS$_2$ are recorded at each point of the $\sim 15\times15\;\mu m^2$ map using an Andor spectrometer. The scan step size is set to $500\;nm$ to ensure spatial overlap of the laser spot between adjacent steps. Due to the high brightness of WS$_2$, we are able to quickly record area scans for different pressure values to obtain a clear view of the behavior of the sample under increasingly non-uniform strain.

\begin{figure}[H]
  \includegraphics[width=\textwidth]{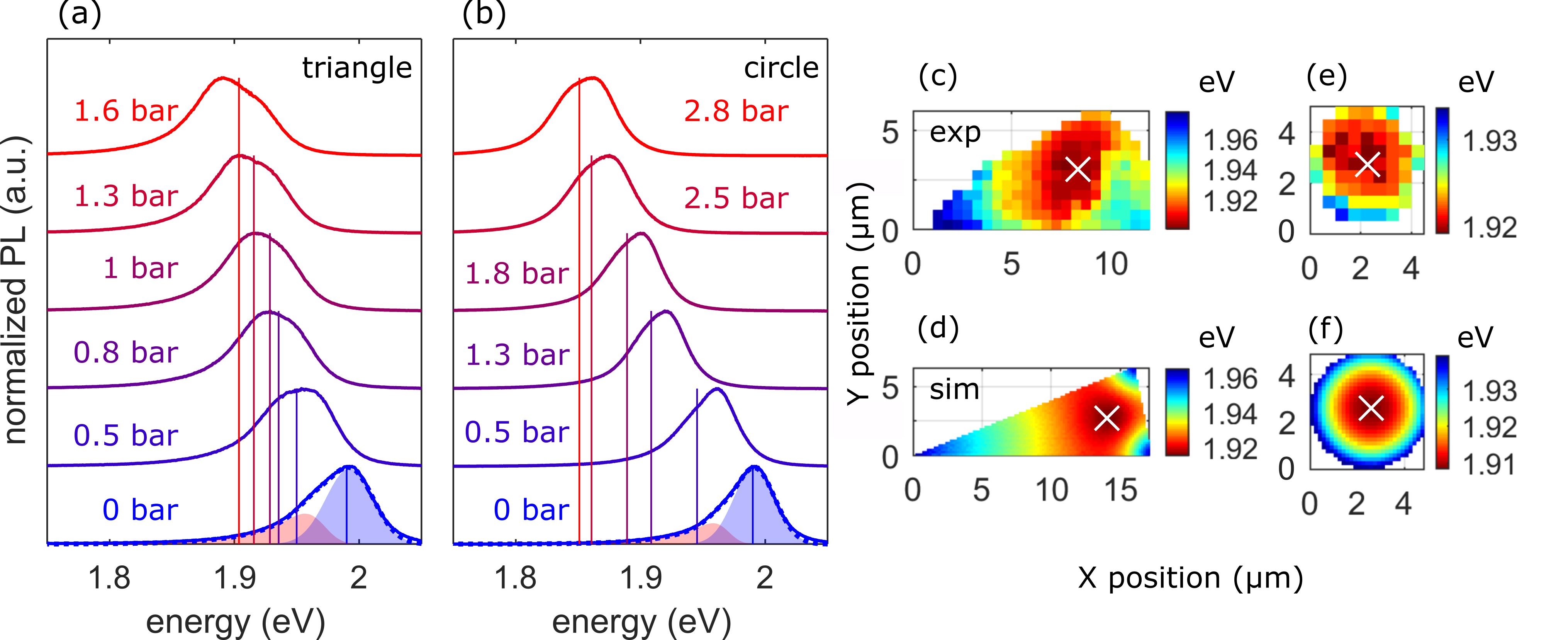}
  \caption{(a-b) PL spectra acquired at the point of maximum strain in a triangle (a) and a circle (b) sample. The shaded areas shown for the lowest curves in (a-b) are the trion (red) and neutral exciton (blue) fitted peaks. The solid vertical lines present the predicted energy of the center of exciton peak position (see main text). The discrepancy between the predicted exciton energy and the peak position is clear for both samples. (c,e) Spatial map of the PL peak maximum at $\sim$ 1.2 bar pressure for a triangular and circular hole, respectively. (d,f) FEM simulations corresponding to (c) and (f). Points of the maximum strain are marked with a cross. Colorbars in (c-f) 
  represent the energy ($eV$). 
  }
  \label{fig:two}
\end{figure}


\section{Results}

Experimentally recorded PL spectra are strongly influenced by applied strain for both triangular and circular samples. We compare the spectra acquired at the point of the highest strain determined via a procedure discussed below. At zero applied pressure, a feature with a maximum at $\sim2\;eV$ is seen for both samples (blue curves in Fig. \ref{fig:two}a,b). The feature is fitted by two peaks, a Gaussian and an exponentially modified Gaussian\citep{recoil}, separated by approximately $30\;meV$ (shaded red and blue curves in Fig. \ref{fig:two}a,b). Two peaks with similar separation are commonly seen in the PL spectra of WS$_2$ and ascribed to neutral excitons (blue-shifted peak) and charged excitons (trions, red-shifted peak)\citep{tri_ex_sep}. The separation between the peaks, $30\;meV$, corresponds to the trion binding energy\citep{tri_ex_sep}.

Two effects are evident in Fig. \ref{fig:two}a,b with increasing pressure: the red-shifting of the spectrum and the modification of the spectral shape. The red-shifting is expected \citep{bunch} and corresponds to strain-related reduction of the band gap of WS$_2$. Following a previous study \citep{bunch}, we extract the spectral position of the PL peak maximum at the point of highest strain (white crosses in Fig. \ref{fig:two}c,e). Assuming that the local band-gap corresponds to the local strain via $E_g(x,y)=E_0-(\varepsilon_{xx}(x,y)+\varepsilon_{yy}(x,y)) \times 50\;meV/\%$\citep{bunch} ($E_0$ is the band-gap at zero strain, $\varepsilon_{xx,yy}(x,y)$ are the spatially-dependent strain tensor components), we calculate the expected spectral position of the maximum PL (see Supporting Information). These are shown as vertical lines for each pressure in Fig. \ref{fig:two}a,b. There is a slight discrepancy between the predicted and measured maximum PL position which will be discussed below.

In addition, we spatially map the experimental maximum PL peak to visualize the distribution of strain (Fig. \ref{fig:two}c,e; maps for other pressures are shown in Supporting Information Fig. S2). We find that in both circular and triangular samples the maximum strain is reached at a point marked by a white cross in Fig. \ref{fig:two}c,e near the middle of the sample. We compare the experimental maps to FEM simulations (Fig. \ref{fig:two}d,f) and find that the strain distribution achieved by the maximum PL peak is comparable to the simulations. This allows us to estimate the non-uniformity of the strain. From the data of Fig. \ref{fig:two}c,e we extract (at $\Delta P=1.7\;bar$) a strain gradient of $\sim 0.1\;\%/\mu m$ for the circular and $\sim 0.28\;\%/\mu m$ for the triangular sample. We achieved maximum strain of $3\%$ for the circular sample and $2\%$ for the triangular sample. Overall, we see that strongly non-uniform strain is induced in both samples and that the degree of non-uniformity is significantly higher for triangular geometry. 

\begin{figure}
  \includegraphics[width=\textwidth]{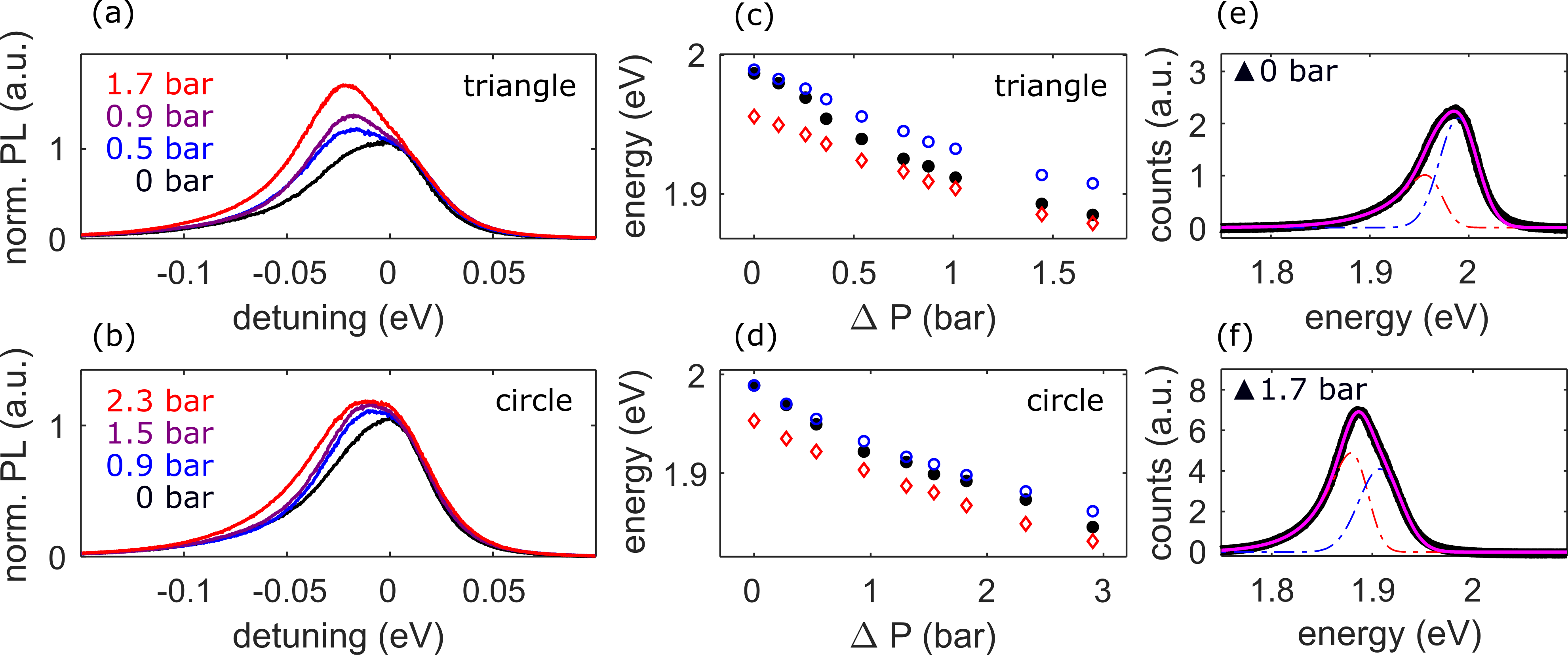}
  \caption{(a,b) Detuning plots, with the neutral exciton peak position centered at $0\;eV$, for a triangle (a) and a circle (b) samples. (c,d) The spectral position of the PL peak maximum (filled black circles), the neutral exciton position (empty blue circles), and the trion position (empty red diamonds) vs. applied pressure for the same samples as in a) and b). The transition from neutral exciton to trion-dominated regimes is more evident for the triangle sample.  (e,f) Examples of experimental spectra at two different pressures (black solid curves) fitted to a neutral exciton (blue dashed curves) and a trion (red dashed curves) resulting in a close fit to the data (pink curve). In (e) the exciton is more dominant while in (f) the trion is more dominant. The asymmetric red tail of the  trion is due to the electron recoil effect.}
  \label{fig:three}
\end{figure}

As the pressure-related spectral red-shifts are well understood, we now examine a more puzzling aspect of our data: pressure-related changes of the spectral shape. Such changes are in stark contrast to uniformly strained TMDCs, where the spectral shape has been shown to be mostly strain-independent\citep{detuning}. To visualize the effect, we plot the spectra for different pressures for both triangular (Fig. \ref{fig:three}a) and circular (Fig. \ref{fig:three}b) samples. The spectra at different pressures were shifted to account for the bandgap red-shift. We see that at high pressures the emission line changes its shape and becomes asymmetric. This effect is much stronger for triangles (Fig. \ref{fig:three}a). 

We note another interesting consequence of the pressure-related peak shape change.
In Fig. \ref{fig:three}c,d we plot both the spectral position of the maximum of the PL peak (black symbols), the neutral exciton position (blue symbols), and the trion position (red symbols) as a function of pressure. For both samples, the neutral exciton position scales linearly with pressure, with a slope of $\sim0.07\;eV/bar$. This is consistent with the expectation from the theory of pressurized membranes, as maximum strain scales linearly with the pressure\citep{membranes_th, vella} and the reduction of the band-gap is a linear function of strain. Less expected is a gradual shift of the PL peak maximum from the position of the neutral exciton to that of the trion. This shift leads to a non-linear pressure dependence of PL peak maximum. Once again, the non-linearity is more pronounced for triangles.

To understand the data of Figures \ref{fig:two} and \ref{fig:three}, we analyze pressure-dependent contributions of neutral excitons and trions to the PL peak. The spectral weight of the trion peak grows while the weight of the neutral exciton peak decreases with increasing pressure (Fig. \ref{fig:three}e,f). To visualize this growth, we plot the ratio between the amplitudes of the trion and neutral excitons peaks obtained from the fitting as a function of $\Delta P$ (Fig. \ref{fig:four}a). This ratio increases with pressure in a roughly linear fashion with larger increase for the triangular sample (red points in Fig. \ref{fig:four}a) than for the circular (blue points in Fig. \ref{fig:four}a).

All of the features presented by the data in Figures \ref{fig:two} and \ref{fig:three} can be explained by assuming the pressure-dependent transfer of the spectral weight from the neutral exciton to trion peak. For example, the gradually increased asymmetry of the spectral lines seen in Fig. \ref{fig:three}a,b reflects the gradual increase of the trion contribution. The observed shift of the maximum PL peak from the neutral exciton peak position at low pressures to the trion peak position at high pressures (Fig. \ref{fig:three}c,d) is also caused by the trion increase. Finally, the shift from neutral exciton to trion-dominated regime  explains the discrepancy between the predicted (vertical lines in Fig. \ref{fig:two}a,b) and the observed spectral position of the maximum PL peak.


 
\section{Discussion}
\begin{figure}
  \includegraphics[width=\textwidth]{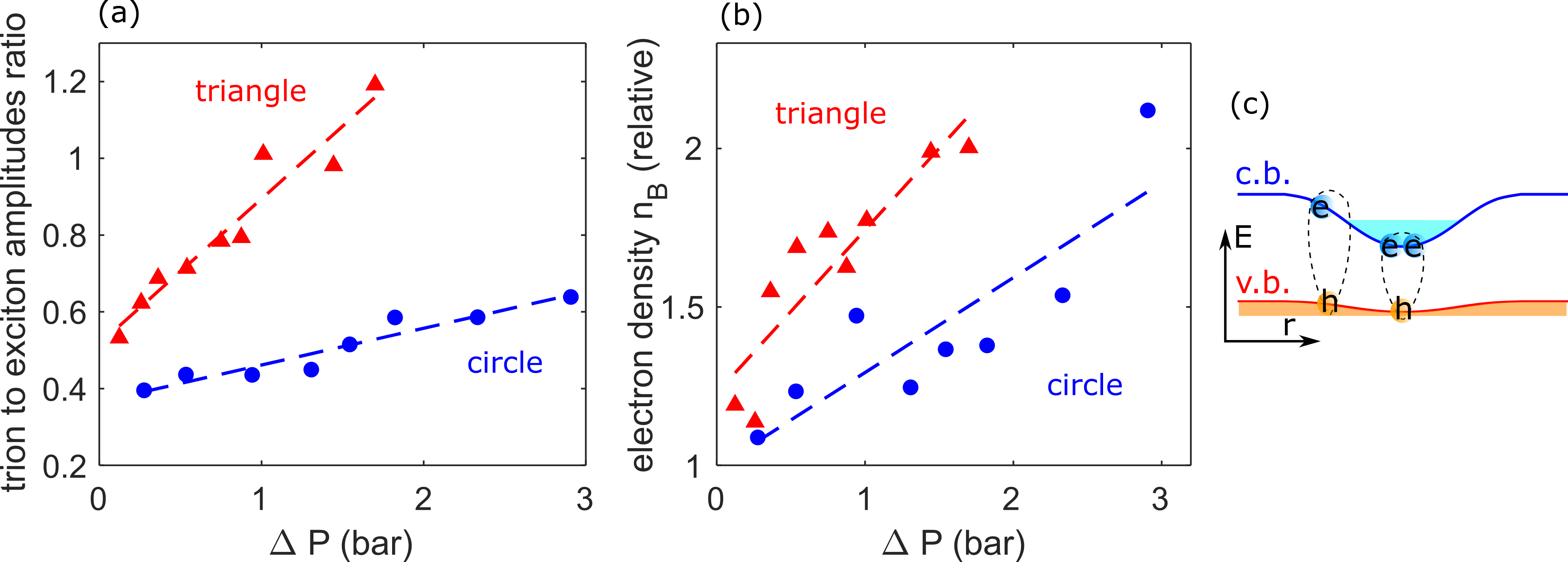}
  \caption{(a) The ratio between the neutral exciton and trion PL peak amplitudes for different samples. The guide-to-the-eye dashed lines show the approximate linear dependency of the conversion efficiency in different samples. (b) Calculation of the free carrier density from the experimental data, see main text. Dashed lines are guide-to-the-eye linear fits. (c) Schematic illustration of the spatial electron funneling under strain-modulated band-gap. }
  \label{fig:four}
\end{figure}

So far, we have shown that mechanical strain has two effects in our samples: the expected red-shift of the PL spectra as well as the unexpected spectral shape changes of the PL peak attributed to the changes in equilibrium between neutral and charged excitons. The following questions remain. Why does the spectral weight of trions in PL spectra from the center of the sample increases with increasing pressure (Fig. \ref{fig:four}a)? Why is this increase larger for the samples in triangular geometry? To answer these questions, we consider the spatial profile  of the band-gap in our devices (Fig. \ref{fig:four}c). Density functional calculations indicate that both conduction and valence bands downshift with increasing strain\citep{strain_pl6, bandgap_dft}. This downshift is maximum at the point of the highest strain in the middle of the device. Critically, free electrons are always present in our devices due to spurious doping and presence of impurities\citep{vacancies}. While at zero pressure these electrons are uniformly distributed throughout the sample, under non-uniform strain profile such as in Fig. \ref{fig:four}c, the electrons are `funneled' towards the point of the lowest conduction band edge at the center of the device\citep{moshe}. Consequently, the density of free electrons in the central region grows with increasing pressure.

Quantitatively, the density of electrons in a spatially-varying potential under thermal equilibrium $n_B(r)$ is obtained from the Boltzmann distribution\citep{moshe, theory1, theory2}:
\begin{equation}
    n_B(r) = \frac{N_0 e^{-\Delta u_c(r)/k_BT}}{\int e^{-\Delta u_c(r)/k_BT} rdr}
    \label{eq:one}
\end{equation}
Here $N_0$ is the density of free carriers in the excitation spot, $\Delta u_c(r)$ is absolute value of the change of the conduction band edge, and $k_BT$ is the thermal energy at room temperature. 

We see from Eq. \ref{eq:one} that the integral in the denominator is dependent on the conduction band spatial change, meaning that if the conduction band gradient is larger in one sample compared to the other, then the free-electron density will be larger. Once the free electron density increases, the trion density increases as well\cite{theory1} as the system converts neutral exciton to trions efficiently.

We used Eq. \ref{eq:one} to determine the density of free electrons at the point of the highest strain for both circular and triangular samples (Fig. \ref{fig:four}b). To accomplish this, we determined the position of the conduction band $\Delta u_c(r)$ directly from experimentally measured PL maps (Fig. \ref{fig:two}c,d), as the spectrum peak red-shifts compared to the spectrum at zero pressure difference. The numerator in Eq. \ref{eq:one} is calculated for the point of the highest strain, and $N_0$ was used as a normalization constant. The carrier density  obtained from numerical integration of Eq. \ref{eq:one} is presented in Fig. \ref{fig:four}b. 

We see that $n_B$ increases roughly linearly with pressure and that the rate of the increase is lower for circles (blue symbols) than for triangles (red symbols). As the density of free electrons grows, neutral excitons capture electrons and convert to trions more efficiently. Quantitatively, it can be shown that in the regime of low excitation powers the ratio between charge and neutral excitons peak amplitudes is proportional to electron density\citep{moshe, theory1, theory2}. Therefore,  a pressure-related increase of $n_B$ by about a factor of two seen in Fig. \ref{fig:four}b for triangles explains a similar increase in the trion peak weight in Fig. \ref{fig:four}a. A comparatively smaller increases in $n_B$ for circles is consistent with weaker pressure-related increase of trions in these samples. We therefore conclude that the spectral changes seen in our non-uniformly strained samples (Figs. \ref{fig:two} and \ref{fig:three}) can all be ascribed to funneling of free electrons to the point of the highest strain and resulting conversion of neutral to charged excitons.

So far we have not mentioned another effect that could potentially arise under non-uniform strain -- funneling of the excitons themselves to the point of maximum strain\citep{funneling}. This effect has been shown to be vanishingly small even in a system with higher absolute strain and higher strain gradient\citep{moshe}. Therefore, in our case, exciton and trion funneling are negligible as well. This justifies the simplifying assumption implied in the previous analysis, that excitons emit from the same point where they are excited. 

We note that in previous work that studied pressurized MoS$_2$ membranes\citep{bunch}, the effect of exciton-to-trion conversion was not observed. This could be due to the fact that with the circular holes which were exclusively considered in that work, smaller strain non-uniformity compared to triangles (as shown in this paper) leads to less prominent conversion to trions. In addition, MoS$_2$ samples used in Ref. \citenum{bunch} are not as bright compared to WS$_2$ and it was consequently more challenging to detect the change of the spectrum.

To summarize, we demonstrated a simple and scalable approach to non-uniform strain engineering of 2D materials. The large size of our samples, as well as high strain amplitude, allows us to visualize strain maps directly and to compare them to FEM simulations. The strain non-uniformity is found to be much higher for triangularly-shaped compared to circularly-shaped samples. Moreover, we confirmed that non-uniform strain produces a unique signature -- a relative increase of the trion peak at the point of the highest strain. We interpret this effect as a result of strain-related funneling of free electrons to that point.

Our results suggest that a minor non-uniform strain component was present in many  earlier experiments dealing with uniform strain engineering. The effects reported here may account for deviations in the data reported there from simple expectations of uniform strain engineering. In the future, it will be especially interesting to investigate the dynamics of the process of neutral-to-charged exciton conversion. Finally, the technique reported here may pave the way towards controlled generation and study of pseudomagnetic fields predicted to arise in non-uniformly strained TMDCs\citep{pseudo_magnetic_tmdc, pseudo_magnetic_tmdc2}.

\begin{acknowledgement}

The authors thank Ben Weintrub, Kirill Greben, Sebastian Heeg and Mengxiong Qiao for experimental help and discussions. This work was supported under ERC Grant No. 639739 and DFG TRR 227.

\end{acknowledgement}


\bibliography{triangles-strain-paper}

\end{document}